\documentclass{elsart}
\journal{Nuclear Physics A}
\usepackage{amssymb}
\usepackage[german,english]{babel}
\usepackage[intlimits,sumlimits,namelimits]{amsmath}
\usepackage[dvips]{epsfig}
\usepackage[dvips]{color}
\usepackage{amssymb,ifthen,shadow}
\usepackage{fancyhdr}
\usepackage[dvips]{epsfig}

\newcommand{\beq}{\begin{equation}}
\newcommand{\eeq}{\end{equation}}
\newcommand{\beqa}{\begin{eqnarray}}
\newcommand{\eeqa}{\end{eqnarray}}
\newcommand{\bra}[1]{\mbox{$\langle #1|$}}
\newcommand{\ket}[1]{\mbox{$|#1\rangle$}}

\begin{document}

\begin{frontmatter}

\title{\bf The $NN$ final-state interaction in the helicity structure
of $\vec d(\vec\gamma,\pi^-)pp$ reaction} 

\author{Eed M.\ Darwish\corauthref{eed}}
\corauth[eed]{{\it E-mail address:} eeddarwish@yahoo.com (E.M.\
Darwish).}

\address{Physics Department, Faculty of Science,
South Valley University,\\ Sohag 82524, Egypt}

\date{\today}

\begin{abstract}
The influence of final-state $NN$-rescattering on the helicity
  structure of the $\vec\gamma\vec d\to\pi^-pp$ reaction in the energy
  range from $\pi$-threshold up to 550 MeV has been investigated. The 
  differential polarized cross-section difference for the parallel and
  antiparallel helicity states is predicted and compared with recent
  experimental data. It is shown that the effect of $NN$-rescattering
  is much less important in the polarized differential cross-section
  difference than in the previously studied unpolarized differential
  cross section. Furthermore, the contribution of $\vec\gamma\vec
  d\to\pi^-pp$ to the spin asymmetry of the deuteron is
  explicitly evaluated over the region of the $\Delta$(1232)-resonance
  with inclusion of $NN$-rescattering. The effect of $NN$ final-state
  interaction is found to be much larger in 
  the asymmetry than in the total cross section and leads to an
  appreciable reduction of the spin asymmetry in the $\Delta$-region.

\vspace{0.2cm}

\end{abstract}
\end{frontmatter}
\section{Introduction}\label{sec1}
A still very interesting topic in intermediate energy nuclear physics
is concerned with the quasifree pion production reaction in
nuclei which is governed by three main mechanisms: (i) The elementary
amplitudes of the four pion production channels possible on the
nucleon, (ii) The Fermi motion of the proton and neutron inside the
nucleus, and (iii) The interaction between the final-state
hadrons. The investigation of pion photo- and electroproduction has
the potential to become an important aspect in meson physics since
many important features of the electromagnetic and hadronic reactions
can be studied through these processes. Interest in this topic has
increased mainly through the construction of new high-duty continuous
electron beam machine such as MAMI in Mainz or ELSA in Bonn.  

The particular interest in pion photoproduction on the deuteron lies
in the fact that the simple and well known deuteron structure allows
one to obtain information on the production process on the neutron
which otherwise is difficult to obtain in view of the absence of any
free neutron targets. The earliest calculations for pion
photoproduction on the deuteron were performed using the impulse
approximation (IA) \cite{ChL51,LaF52}. Approximate treatments of
final-state interaction (FSI) effects within a diagrammatic approach
have been reported in \cite{BlL77,Lag78,Lag81}. The authors noted that
the FSI effects are quite small for the charged-pion production
channels in comparison to the neutral one. Photoproduction of pions on
the deuteron has been investigated in the spectator nucleon model
\cite{ScA96} neglecting all kinds of FSI and two-body processes. The
$NN$-FSI has been considered in \cite{Lev01} and good agreement with
experiment was achieved. The influence of final-state $NN$- and $\pi
N$-rescattering on the unpolarized cross sections has been
investigated in \cite{Dar03}. It has been found that $\pi
N$-rescattering is much less important (in general negligible)
compared to $NN$-rescattering. Inclusion of such effects leads to good
agreement with experiment. The role of the $N\Delta$-FSI in the pion
photoproduction off the deuteron has been investigated in
\cite{Faes02}. It has been shown that full calculations with the
off-shell amplitudes of $NN$- and $N\Delta$-FSI are necessary to
obtain a quantitative description of the cross sections.

Up to now, most of calculations have considered only the unpolarized
observables like differential and total cross sections. These cross
sections provide information only on the sum of the absolute squares
of the amplitudes, whereas polarization observables allow extraction
of more information. Observables with polarized photon beam and/or
polarized deuteron target have been poorly investigated. The
particular interest in these observables is based on the fact that, a
series measurements of the polarization observables in photoproduction
reactions have been carried out or planned at different laboratories.
The GDH collaboration has undertaken a joint effort towards the
experimental verification of the Gerasimov-Drell-Hearn (GDH) sum rule,
measuring the difference of the helicity components in total and
differential photoabsorption cross sections. Our goal is to analyze
these experimental measurements.

Recently, polarization observables for incoherent pion photoproduction
on the deuteron have been studied in
\cite{Log00,Dar03+,Dar04,Dar04E13,Dar05JG,Log04}. $\pi^-$-production
channel has been studied within a diagrammatic approach \cite{Log00}
including $NN$- and $\pi N$-rescattering. In that work, predictions
for analyzing powers connected to beam and target polarization, and to
polarization of one of the final protons are presented. In our
previous evaluation \cite{Dar03+}, special emphasize is given for the
beam-target spin asymmetry and the GDH sum rule. Single- and
double-spin asymmetries for incoherent pion photoproduction on the
deuteron have been predicted in \cite{Dar04,Dar04E13,Dar05JG} without
any kind of FSI effects. The target tensor analyzing powers of the
$d(\gamma,\pi^-)pp$ reaction have been studied in the plane wave
impulse approximation \cite{Log04}. Most recently, our evaluation
\cite{Dar03+} has been extended to higher energies in \cite{Aren04}
with additional inclusion of two-pion and eta production.

As a further step in this study, we investigate in this paper the
influence of $NN$ FSI effect on the polarized differential and total
cross sections with respect to parallel and antiparallel spins of
photon and deuteron for $\gamma d\to\pi^-pp$. Our second point of
interest is to analyze the recent experimental data from the GDH
collaboration \cite{Pedroni}. With respect to the interactions in the
final two-body subsystems, only the $NN$-rescattering is obtained into
account since $\pi N$-rescattering has been considered as negligible 
\cite{Lev01,Dar03}. 

In Sect.\ \ref{sec2} of this paper, the model for the elementary
$\gamma N\to\pi N$ and $NN\to NN$ reactions which will serve as an
input for the reaction on the deuteron is briefly reviewed. Sect.\ 
\ref{sec3} will introduce the general formalism for incoherent pion
photoproduction on the deuteron. The separate contributions of the IA
and the $NN$-rescattering to the transition matrix are described in
this section. Details of the actual calculation and the results are
presented and discussed in Sect.\ \ref{sec5}. Finally, a summary and
conclusions are given in Sect.\ \ref{sec6}.
\section{The elementary $\gamma N\to\pi N$ and $NN\to NN$
reactions}\label{sec2} 
Pion photoproduction on the deuteron is governed by basic 
two-body processes, namely pion photoproduction on a nucleon and 
hadronic two-body scattering reactions. For the latter only
nucleon-nucleon scattering is considered in
this work. As already mentioned in the introduction, $\pi
N$-rescattering is found to be negligible and thus it is not
considered in the present calculation. 

The starting point of the construction of an operator for
pion photoproduction on the two-nucleon space is the elementary pion
photoproduction operator on a single nucleon, i.e., $\gamma N\to\pi
N$. In the present work we  will examine the various observables for
pion photoproduction on the free nucleon using, as in our previous
work \cite{Dar03}, the effective Lagrangian model developed by Schmidt
{\it et al.}  \cite{ScA96}. The main advantage of this model is that
it has been constructed to give a realistic description of the
$\Delta$(1232)-resonance region. It is also given in an  arbitrary
frame of reference and allows a well defined off-shell continuation as
required for studying pion production on nuclei. This model consists
of the standard pseudovector Born terms and the contribution of the
$\Delta(1232)$-resonance. For further details with respect to the
elementary pion photoproduction operator we refer to \cite{ScA96}. As
shown in Figs.\ 1-3 in our previous work \cite{Dar03}, the results of
our calculations for the elementary process are in good agreement with
recent experimental data as well as with other theoretical predictions
and gave a clear indication that this elementary operator is quite
satisfactory for our purpose, namely to incorporate it into the
reaction on the deuteron.  

For the nucleon-nucleon scattering in the $NN$-subsystem we use in
this work a specific class of separable potentials \cite{HaP8485}
which historically have played and still play a major role in the
development of few-body physics and also fit the phase shift data for
$NN$-scattering. The EST method \cite{Ern7374} for constructing
separable representations of modern $NN$ potentials has been  applied
by the Graz group \cite{HaP8485} to cast the  Paris potential
\cite{La+80} in separable form. This separable model is most widely
used in case of the $\pi NN$ system (see for example \cite{Gar90} and
references therein). Therefore, for the present study of the influence
of $NN$-rescattering this model is good enough.
\section{$\pi$-photoproduction off the deuteron}\label{sec3}
The formalism of incoherent pion photoproduction on the deuteron is
presented in detail in our previous work \cite{Dar03}. Here we briefly
recall the necessary notation and definitions. The general expression
for the unpolarized cross section according to \cite{BjD64} is given
by 
\begin{eqnarray}
d\sigma &=&
\frac{\delta^{4}(k+d-p_1-p_2-q)M^2_{N}d^{3} p_1d^{3} p_2d^{3} q}
{96(2 \pi)^{5} |\vec v_{\gamma} - \vec
v_{d}|\omega_{\gamma}E_{d}E_1E_2\omega_q} \nonumber \\
& & \times~ \sum_{s\,m\,t\,,m_{\gamma}\,m_d} 
\left |{\mathcal M}^{(t\,\mu)}_{s\,m\,m_{\gamma}\,m_d}
(\vec p_1,\vec p_2,\vec q,\vec k,\vec d) \right|^{2}\,,
\label{eq:3.2}
\end{eqnarray}
where $k=(\omega_{\gamma},\vec k)$, $d=(E_d,\vec d)$,
$q=(\omega_q,\vec q)$, $p_1=(E_1,\vec p_1)$ and $p_2=(E_2,\vec p_2)$
denote the 4-momenta of photon, deuteron, pion and two nucleons,
respectively. Furthermore, $m_{\gamma}$ denotes the photon
polarization, $m_{d}$ the spin projection of the deuteron, $s$ and $m$
total spin and projection of the two outgoing nucleons, respectively,
$t$ their total isospin, $\mu$ the isospin projection of the pion, and
$\vec{v}_{\gamma}$ and $\vec{v}_{d}$ the velocities of photon and
deuteron, respectively. The transition amplitude is denoted by
${\mathcal M}$. Covariant state normalization in the 
convention of \cite{BjD64} is assumed. 

\begin{figure}[htb]
\includegraphics[scale=1.0]{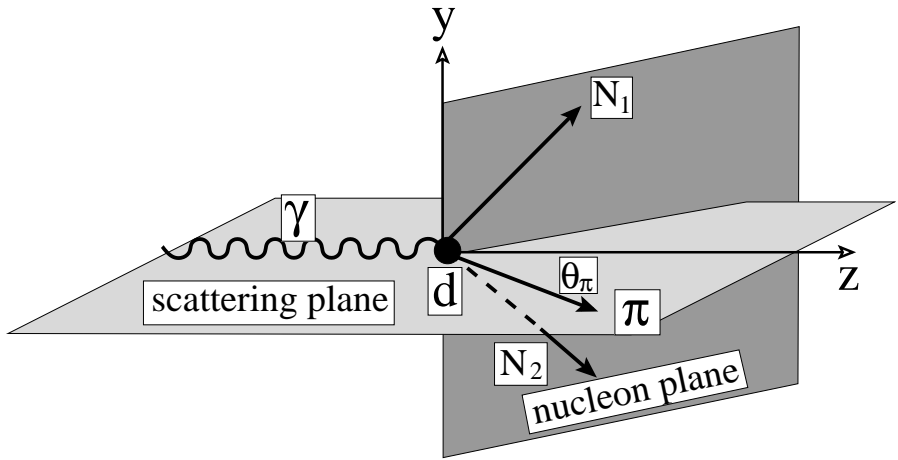}
  \caption{\small Kinematics in the laboratory system for pion
  photoproduction on the deuteron.}
  \label{labsys}
\end{figure}

This expression is evaluated in the lab or deuteron rest frame. 
A right-handed coordinate system is chosen where the
$z$-axis is defined by the photon momentum $\vec k$ and the
$y$-axis by $\vec k \times \vec q$. The scattering
plane is defined by the momenta of photon $\vec k$ and pion  
$\vec q$ whereas the momenta of outgoing nucleons  
$\vec p_1$ and $\vec p_2$ define the nucleon plane (see Fig.\ 
\ref{labsys}). As independent
variables, the pion momentum $q$, its angles $\theta_{\pi}$
and $\phi_{\pi}$, the polar angle $\theta_{p_{NN}}$ and the azimuthal
angle $\phi_{p_{NN}}$ of the relative momentum $\vec p_{NN}$ of the
two outgoing nucleons are chosen. The total and relative momenta of
the final $NN$-system are defined by $\vec{P}_{NN} = \vec{p}_{1} +
\vec{p}_{2}= \vec{k} - \vec{q}$ and $\vec p_{NN} =
\frac{1}{2}\left(\vec{p}_{1} - \vec{p}_{2}\right)$, respectively.

Integrating over the pion momentum $q$ and over $\Omega_{p_{NN}}$, one
obtains the semi-inclusive differential cross section of pion
photoproduction on the deuteron, where only the final pion is detected
without analyzing its energy,  
\beq
\frac{d\sigma}{d\Omega_{\pi}}=\int_0^{q_{max}}dq
\int d\Omega_{p_{NN}}\,
\frac{\rho_{s}}{6}\,\sum_{s\,m\,t\,m_{\gamma}\,m_{d}}  
\left \vert {\mathcal
M}^{(t\,\mu)}_{s\,m\,m_{\gamma}\,m_{d}}(\vec{p}_{NN},\vec q,\vec k)
\right \vert^{2}\,, 
\eeq
where $\rho_{s}$ denotes the phase space factor (see Eq.\ (7) in 
\cite{Dar03} for its definition).

\begin{figure}[htb]
\hspace*{1cm}\includegraphics[scale=.8]{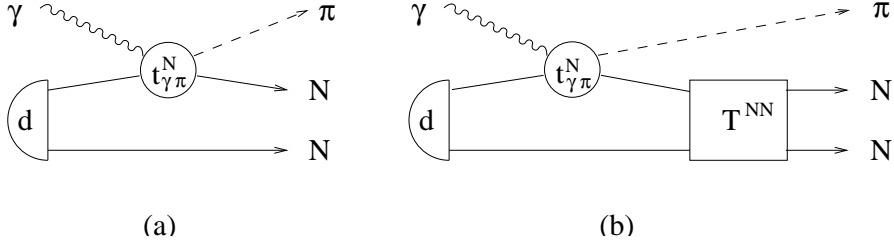}
\caption{Diagramatic representation of pion photoproduction on the
deuteron including $NN$-rescattering in the final state: (a) impulse
approximation (IA) and (b) $NN$-rescattering.} 
\label{t-matrix}
\end{figure}

The general form of the photoproduction transition matrix is given by
\begin{eqnarray}\label{general}
{\mathcal M}^{(t\mu)}_{sm
m_{\gamma}m_d}(\vec{k},\vec{q},\vec{p_1},\vec{p_2}) 
& = &
^{(-)}\bra{\vec{q}\,\mu,\vec{p_1}\vec{p_2}\,s\,m\,t-\mu}\epsilon_{\mu}  
(m_{\gamma})J^{\mu}(0)\ket{\vec{d}\,m_d\,00}\,,
\end{eqnarray}
where $J^{\mu}(0)$ denotes the current operator. The outgoing $\pi NN$
scattering state is approximated in this 
work by  
\begin{eqnarray}
\ket{\vec{q}\,\mu,\vec{p_1}\vec{p_2}\,s\,m\,t-\mu}^{(-)} &=&
\ket{\vec{q}\,\mu,\vec{p_1}\vec{p_2}\,s\,m\,t-\mu} \nonumber \\
& & + ~G_{0}^{\pi NN (-)}
\,T^{NN} \ket{\vec{q}\,\mu,\vec{p_1}\vec{p_2}\,s\,m\,t-\mu}\,,
\end{eqnarray}
where $\ket{\vec{q}\,\mu,\vec{p_1}\vec{p_2}\,s\,m\,t-\mu}$ denotes the 
free $\pi NN$ plane wave, $G_{0}^{\pi NN (-)}$ the free $\pi NN$
propagator, and $T^{NN}$ the reaction operator for $NN$-scattering. Thus,
the total transition matrix element reads in this approximation  
\begin{eqnarray}
\label{threethree}
{\mathcal M}^{(t\mu)}_{sm m_{\gamma}m_d} & = &
{\mathcal M}_{sm m_{\gamma}m_d}^{(t\mu)~IA} + 
{\mathcal M}_{sm m_{\gamma}m_d}^{(t\mu)~NN}\,. 
\end{eqnarray}
A graphical representation of the transition matrix is shown in
Fig.~\ref{t-matrix}.

As shown in \cite{Dar03}, the matrix element in the IA has the
following expression  
\begin{eqnarray}\label{g16}
  {\mathcal M}_{sm m_{\gamma}m_d}^{(t\mu)~IA}
  (\vec k,\vec q,\vec p_1,\vec p_2) &=&
 \sqrt{2}\sum_{m^{\prime}}\langle s 
  m,\,t -\mu|\,\Big( \langle
  \vec{p}_{1}|t_{\gamma\pi}(\vec k,\vec q\,)|-\vec{p}_{2}\rangle
  \tilde{\Psi}_{m^{\prime},m_{d}}(\vec{p}_{2}) 
\nonumber\\ & &  \hspace{1cm} 
-(-)^{s+t}(\vec p_1 \leftrightarrow \vec p_2) 
\Big)\,|1
  m^{\prime},\,00\rangle \,,
\end{eqnarray}
where $t_{\gamma\pi}$ denotes the elementary production amplitude on
the nucleon and $\widetilde{\Psi}_{m,m_{d}}(\vec{p}\,)$ is given by 
\begin{equation}
  \widetilde{\Psi}_{m,m_{d}}(\vec{p}\,) =
  (2\pi)^{\frac{3}{2}}\sqrt{2E_{d}}
  \sum_{L=0,2}\sum_{m_{L}}i^{L}\,C^{L 1 1}_{m_{L} m m_{d}}\,
  u_{L}(p)Y_{Lm_{L}}(\hat{p}) \,.
\end{equation}
For the radial deuteron wave function $u_{L}(p)$, the Paris potential
\cite{La+81} is used.

For the $NN$-rescattering contribution, one obtains \cite{Dar03}
\begin{eqnarray}
\label{tnn-fsi-final}
{\mathcal M}^{(t\mu)~NN}_{sm m_{\gamma}m_d}
(\vec k,\vec q,\vec p_1,\vec p_2) & = & 
\sum_{m^{\prime}}\int d^3\vec p^{\,\prime}_{NN} 
\sqrt{\frac{E_1 E_2}{E_1' E_2'}}
\,\widetilde {\mathcal R}_{s m m^{\prime}}^{NN,\,t\mu}(W_{NN},\vec
p_{NN},\vec p^{\,\prime}_{NN}) \nonumber\\
&&\times~\frac{M_N}{\widetilde p^{\, 2} - p_{NN}^{\prime\,2} + i\epsilon}
{\mathcal M}^{(t\mu)~IA}_{sm^{\prime},m_{\gamma}m_d}(\vec k,\vec
q,\vec p^{\,\prime}_1,\vec p^{\,\prime}_2)\,,
\end{eqnarray}
where $\vec p_{NN}^{\,\prime}=\frac{1}{2}\,(\vec{p}_1^{\,\prime}-
\vec{p}_2^{\,\prime})$ denotes the relative momentum of the
interacting nucleons in the intermediate state, $W_{NN}$ is the
invariant mass of the $NN$-subsystem, $\vec
p^{\,\prime}_{1/2}=\pm \vec p^{\,\prime}_{NN} + (\vec k-\vec q\,)/2$
and $E_{1/2}'$ are the momenta and the corresponding on-shell energies
of the two nucleons in the intermediate state, respectively, and
$\widetilde p^{\,2} = M_N(E_{\gamma
d}-\omega_{\pi}-2M_N-(\vec{k}-\vec{q})^2/4M_N)$ with $E_{\gamma
d}=M_d+\omega_{\gamma}$. The conventional $NN$-scattering matrix
$\widetilde {\mathcal 
R}_{smm^{\prime}}^{NN,\,t\mu}$ is introduced with respect to
noncovariantly normalized states. It is expanded in terms of the
partial wave contributions ${\mathcal
T}_{Js\ell\ell^{\prime}}^{NN,\,t\mu}$  as follows 
\begin{eqnarray}
\label{tnn-hos}
\widetilde {\mathcal R}_{smm^{\prime}}^{NN,\,t\mu}(W_{NN},\vec
 p_{NN},\vec p^{\,\prime}_{NN}) & = & \sum_{J\ell\ell^{\prime}}
 {\mathcal F}_{\ell\ell^{\prime}\,mm'}^{NN,\,Js}
 (\hat{p}_{NN},\hat{p}_{NN}^{\,\prime}) \nonumber \\
 & & \times~{\mathcal T}_{Js\ell\ell^{\prime}}^{NN,\,t\mu}
 (W_{NN},p_{NN},p_{NN}^{\,\prime})\, , 
\end{eqnarray}
where the purely angular function ${\mathcal
F}_{\ell\ell^{\prime} mm^{\prime}}^{NN,\,Js}
(\hat{p}_{NN},\hat{p}_{NN}^{\,\prime})$ is defined by  
\begin{eqnarray}
\label{tnn-ff}
{\mathcal F}_{\ell\ell^{\prime} mm^{\prime}}^{NN,\,Js}(\hat{p}_{NN},
\hat{p}_{NN}^{\,\prime}) & = & 
 \sum_{M m_{\ell}m_{\ell^{\prime}}} C^{\ell s J}_{m_{\ell} m M}\,
C^{\ell^{\prime} s J}_{m_{\ell^{\prime}} m^{\prime} M}
 Y^{\star}_{\ell m_{\ell}}(\hat{p}_{NN})
 Y_{\ell^{\prime} m_{\ell^{\prime}}}(\hat{p}_{NN}^{\,\prime})\,.
\end{eqnarray}
The necessary half-off-shell $NN$-scattering matrix ${\mathcal
T}_{Js\ell\ell^{\prime}}^{NN,\,t\mu}$ was obtained from separable
representation of a realistic $NN$-interaction \cite{HaP8485} which
gave a good description of the corresponding phase shifts. Explicitly,
all partial waves with  total angular momentum $J\le 3$ have been
included. 
\section{Results and discussion}\label{sec5}
The discussion of our results is divided into two parts. First, 
we will discuss the influence of $NN$-FSI effect on the
polarized differential cross-section difference
$({d\sigma/d\Omega_{\pi}})^P-({d\sigma/d\Omega_{\pi}})^A$  for the
parallel and antiparallel helicity states by comparing the pure IA
with the  inclusion of $NN$-rescattering in the final
state. Furthermore, we will confront our results with recent
experimental data from the GDH collaboration \cite{Pedroni}. In the
second part, we will then consider the polarized total cross 
sections for circularly polarized photons on a target with spin
parallel $\sigma^P$ and antiparallel $\sigma^A$ to the photon
spin. The contribution of 
$\vec\gamma\vec d\to\pi^-pp$ to the spin 
response of the deuteron, i.e., the asymmetry of the total
photoabsorption cross-section with respect to parallel and
antiparallel spins of photon and deuteron, has been explicitly
evaluated over the range of the $\Delta$(1232)-resonance with
inclusion of final-state $NN$-rescattering.  
\subsection{The helicity difference
$({d\sigma/d\Omega_{\pi}})^P-({d\sigma/d\Omega_{\pi}})^A$}\label{sec51} 
\begin{figure}[htp]
\begin{center}
\includegraphics[scale=0.8]{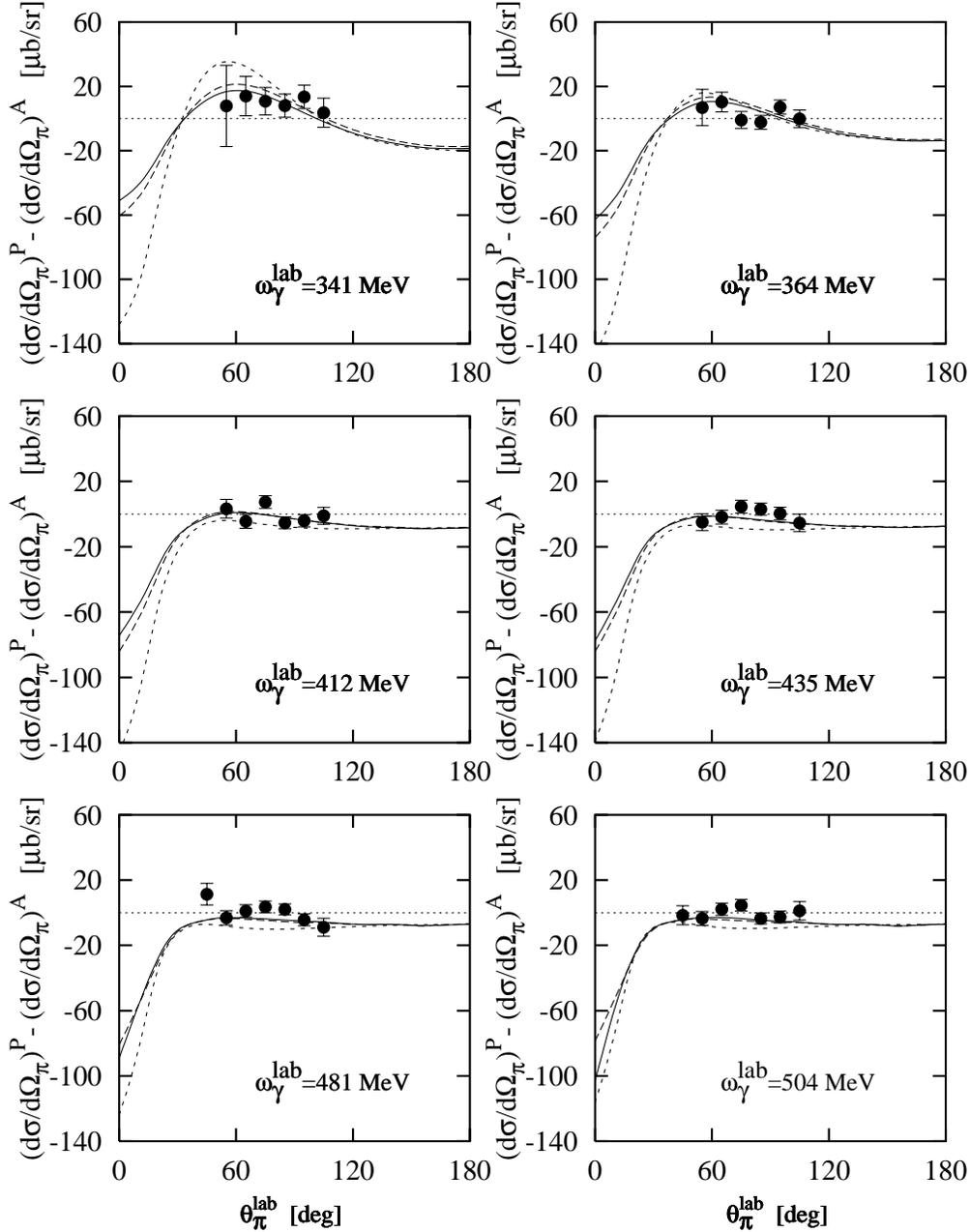}
\caption{The differential polarized cross-section difference
$({d\sigma/d\Omega_{\pi}})^P-({d\sigma/d\Omega_{\pi}})^A$ for
$\vec\gamma\vec d\to\pi^-pp$ for the parallel 
$({d\sigma/d\Omega_{\pi}})^P$ and antiparallel
$({d\sigma/d\Omega_{\pi}})^A$ helicity states as a function of
pion angle in the laboratory frame in comparison to recent measurement
from \cite{Pedroni} at different values of photon
lab-energy. Notation: dashed curves: IA; solid
curves: IA+$NN$-rescattering; dotted curves: predictions for $\pi^-$
production on the free neutron, i.e., $\vec\gamma\vec n\to\pi^-p$.} 
\label{diff}
\end{center}
\end{figure}

\begin{figure}[htp]
\begin{center}
\includegraphics[scale=0.8]{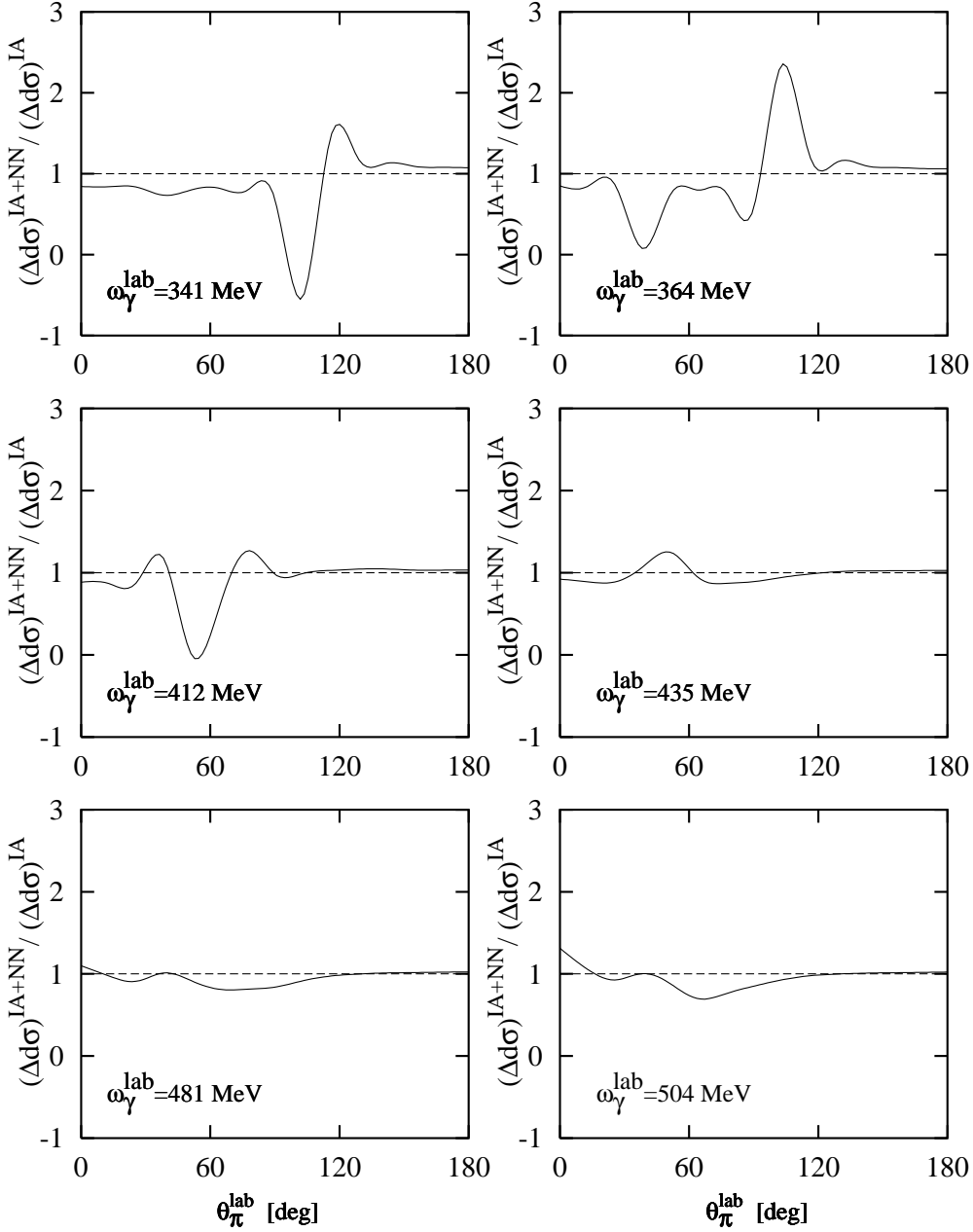}
\caption{The ratio $(\Delta d\sigma)^{IA+NN}/(\Delta d\sigma)^{IA}$ 
(see Eq.\ (\ref{ratio}) for its definition) as a function of pion
angle in the laboratory frame at different photon lab-energies.} 
\label{diffdiff}
\end{center}
\end{figure}

We begin the discussion with presenting our results for the
differential polarized cross-section difference for the parallel
$({d\sigma/d\Omega_{\pi}})^P$ and antiparallel
$({d\sigma/d\Omega_{\pi}})^A$ helicity states in the pure IA and with
$NN$-rescattering  as shown in Fig.\ \ref{diff} as a function of
emission pion angle in the laboratory frame at different values of
photon lab-energy. One readily notes, that $NN$-rescattering - the
difference between the dashed and the solid curves - is quite small,
almost completely negligible at pion backward angles. The reason for
that stems from the fact that in charged-pion production, the
$^3S_1$-contribution to the $NN$ final state is forbidden.  
In order to give a more detailed and quantitative evaluation of
$NN$-FSI on the differential polarized cross-section
difference, we show in Fig.\ \ref{diffdiff} the relative effect by
plotting the ratio of the corresponding cross-section difference to
the ones for the IA, i.e., 
\begin{eqnarray}
\frac{(\Delta d\sigma)^{IA+NN}}{\hspace*{-0.6cm}(\Delta d\sigma)^{IA}}
&~=~&  
\frac{\left[(\frac{d\sigma}{d\Omega_{\pi}})^P -
(\frac{d\sigma}{d\Omega_{\pi}})^A\right]^{IA+NN}}
{\hspace*{-0.6cm}\left[(\frac{d\sigma}{d\Omega_{\pi}})^P -
(\frac{d\sigma}{d\Omega_{\pi}})^A\right]^{IA} } \,.
\label{ratio}
\end{eqnarray}
One sees that the major contribution
from $NN$-FSI appears at forward pion angles. This
contribution is much less important in the differential polarized
cross-section difference than in the previously studied unpolarized
differential cross sections (compare with Fig.\ 13 in
\cite{Dar03}). It has been found that $NN$-FSI reduces the unpolarized 
differential cross section by about 15$\%$ at
$\theta_{\pi}=0^{\circ}$ \cite{Dar03}. This reduction decreasing
rapidly with increasing pion angle. 

By comparing the results of the difference 
$({d\sigma/d\Omega_{\pi}})^P-({d\sigma/d\Omega_{\pi}})^A$ for
$\vec\gamma\vec d\to\pi^-pp$ (solid curves in Fig.\ \ref{diff})
with those for the free $\vec\gamma\vec n\to\pi^-p$ case (dotted
curves in Fig.\ \ref{diff}), we see that a large amount of correction
is needed to go from the bound deuteron to the free neutron 
case. The difference between both results decreases to a tiny effect
at backward angles. Fig.\ \ref{diff} shows also a comparison of our
results for the helicity difference with the experimental data from the
GDH collaboration \cite{Pedroni}. It is obvious that a quite
satisfactory agreement with experiment is achieved. An experimental
check of the helicity difference at extreme forward and backward
pion angles is needed. Also, an independent check in the framework of
effective field theory would be very interesting.
\subsection{Polarized total cross sections}\label{sec52}
\begin{figure}[htp]
\begin{center}
\includegraphics[scale=0.8]{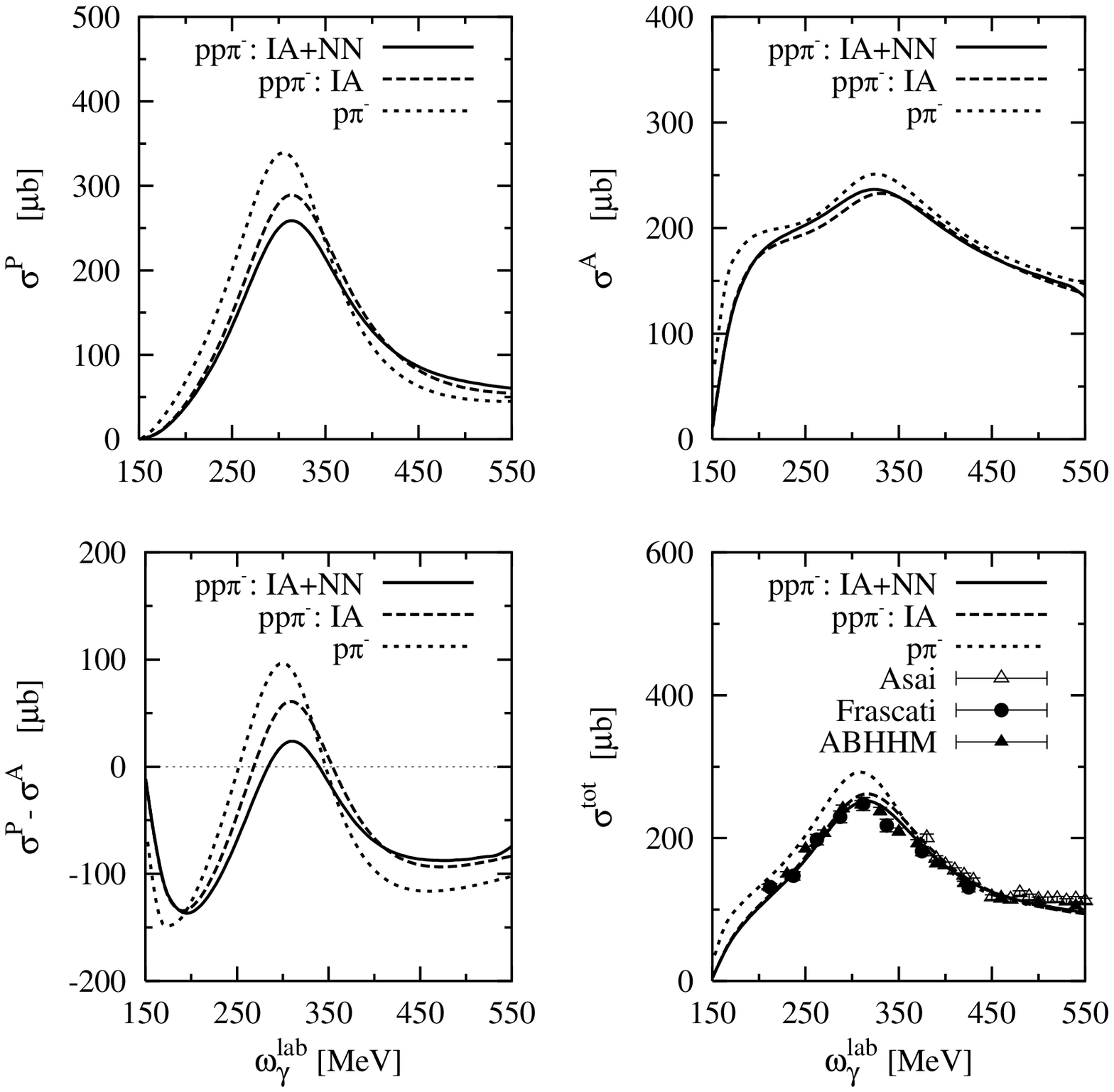}
\caption{The total photoabsorption cross-sections for
circularly polarized photons on a target with spin parallel $\sigma^P$ 
(upper part: left) and antiparallel $\sigma^A$ (upper part: right) to
the photon spin for $\vec\gamma \vec d\to\pi^-pp$ as 
functions of photon lab-energy. The lower part shows the difference
$\sigma^P$-$\sigma^A$ (lower part: left) and the unpolarized total cross
section (lower part: right). The experimental data are 
from \cite{Be+73} (ABHHM), \cite{ChD75} (Frascati) and \cite{As+90}
(Asai). Notation of the curves as in Fig.\ \ref{diff}.}  
\label{total}
\end{center}
\end{figure}

\begin{figure}[htp]
\begin{center}
\includegraphics[scale=0.8]{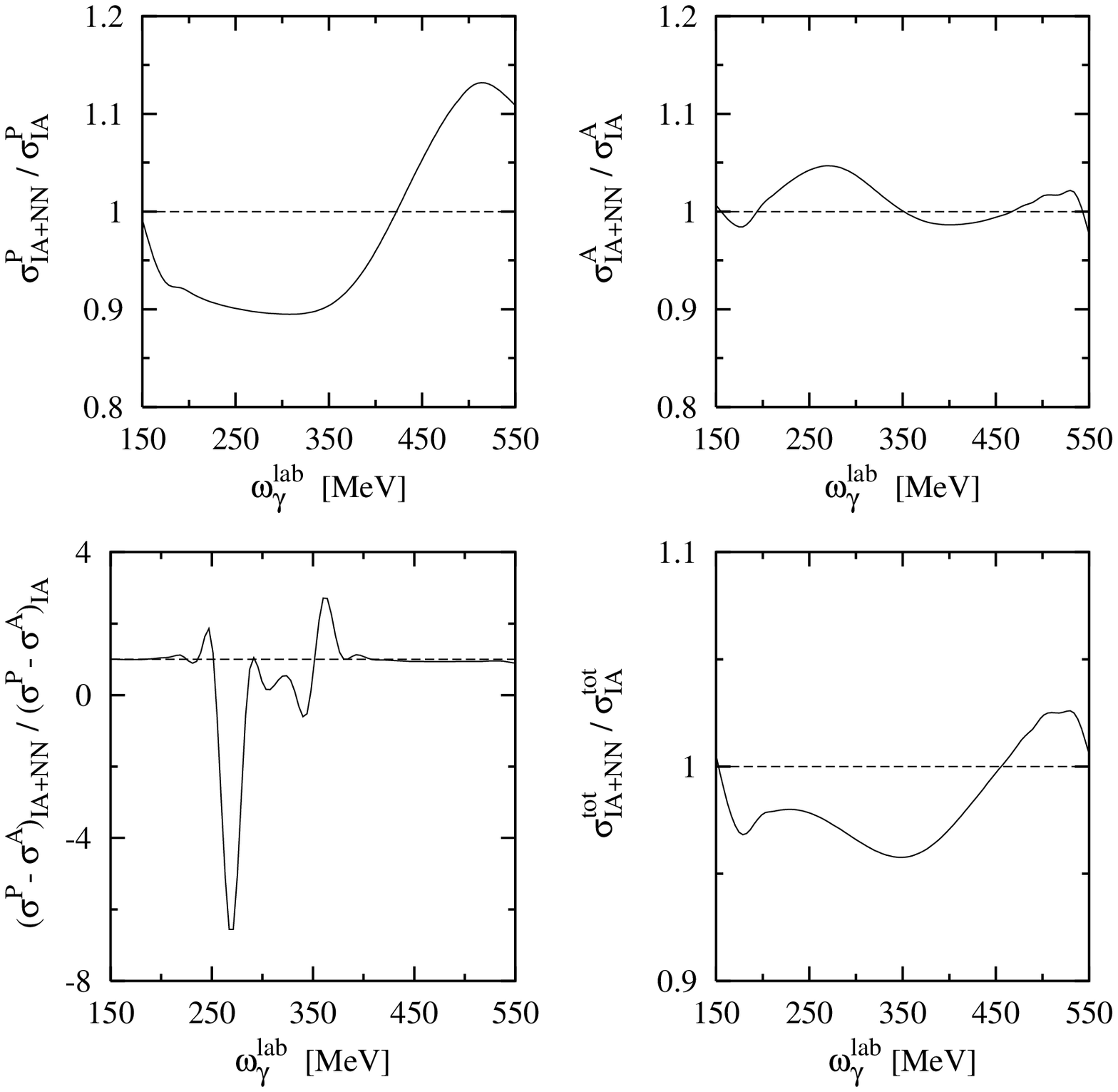}
\caption{The ratios
$\sigma^P_{IA+NN}/\sigma^P_{IA}$ (upper part: left),  
$\sigma^A_{IA+NN}/\sigma^A_{IA}$ (upper part: right),  
$(\sigma^P-\sigma^A)_{IA+NN}/(\sigma^P-\sigma^A)_{IA}$ (lower part:
left) and $\sigma^{tot}_{IA+NN}/\sigma^{tot}_{IA}$ (lower part: right)
as functions of photon lab-energy.} 
\label{totalratio}
\end{center}
\end{figure}

Here the results for the polarized total cross sections in IA alone
and with $NN$-FSI effect are presented as shown in
Fig.\ \ref{total}, where the left top panel shows the total
photoabsorption cross section $\sigma^P$ for circularly polarized
photons on a target with spin parallel to the photon spin, the
right top panel shows the one for antiparallel spins of 
photon and target $\sigma^A$, the left bottom panel shows the spin
asymmetry $\sigma^P-\sigma^A$ and the right bottom panel shows the
results for the unpolarized total cross section in comparison with the
experimental data from \cite{Be+73} (ABHHM), \cite{ChD75} (Frascati) 
and \cite{As+90} (Asai). For comparison, we also show in the same
figure the results for $\pi^-$ production on the free neutron by the
dotted curves. 
In order to see more clearly the relative size of the
interaction effect, we have plotted in Fig.\ \ref{totalratio} the
ratios with respect to the IA.

One notes for the cross sections $\sigma^P$ and $\sigma^A$, the spin
asymmetry $\sigma^P-\sigma^A$ as well as for the unpolarized total 
cross section of the nucleon and the deuteron qualitatively a similar 
behaviour, although for the deuteron the maxima and minima are smaller
and also slightly shifted towards higher energies. Furthermore, in
the case of $\sigma^P$ a large deviation between the IA
and the elementary one - the difference between the dashed and the
dotted curves - is seen because of the Fermi motion and
FSI, whereas for $\sigma^A$ the difference is
smaller. $NN$-FSI effect appears mainly in
$\sigma^P$. 
The left bottom panel in Fig.\ \ref{total} shows that the
helicity difference of the total cross section ($\sigma^P-\sigma^A$)
starts out negative due to the $E_{0+}$ multipole which is dominant in
the threshold region and has a strong positive contribution due to the
$M_{1+}$ multipole which is dominant in the $\Delta$(1232)-resonance
region. It is also clear that FSI leads to a
strong reduction of the spin asymmetry in the energy region of the
$\Delta$(1232)-resonance. This reduction becomes about 35 $\mu$b in
the maximum. Thus, the IA is not a reasonable approximation as it is
for the unpolarized total cross section. Moreover, already the IA
deviates significantly from the corresponding nucleon quantities. It
is also obvious that $\sigma^P$ is much larger than $\sigma^A$ because
of the $\Delta$-excitation. 

For the unpolarized total cross section in the
right bottom panel of Fig.\ \ref{total}, one also notes that $NN$-FSI 
effect is  small, not more than about 5
percent. This effect comes mainly from the change in the radial wave
function of the final $NN$ partial waves by the
interaction. Therefore, it reduces the cross section.  The charged
final state $p\pi^-$ had been investigated 30 years ago in a bubble
chamber measurement of the $\gamma d\to pp\pi^-$ reaction  by the
ABHHM collaboration \cite{Be+73}, at Frascatti \cite{ChD75}, and later
at higher energies by the TAGX-collaboration \cite{As+90}.  The right
bottom panel of Fig.\ \ref{total} shows a comparison between our
results and this set of 
experimental data. One readily notes, that the inclusion of
$NN$-rescattering improves the agreement between experimental data and
theoretical predictions considerably.
\section{Summary and conclusions}\label{sec6}
The influence of $NN$-FSI effect on the polarized
differential and total cross-section differences 
$({d\sigma/d\Omega_{\pi}})^P-({d\sigma/d\Omega_{\pi}})^A$ and
$\sigma^P-\sigma^A$, respectively, for the parallel and antiparallel
helicity states for the $\vec\gamma\vec d\to\pi^- pp$ reaction is
investigated. These helicity asymmetries give valuable information on
the nucleon spin structure and allow a test of the GDH sum rule. For
the elementary production 
operator on the  nucleon, an effective Lagrangian model is used. As
model for the interaction of the $NN$-subsystem we used separable
representation of realistic $NN$ interaction which give a good
description of the corresponding phase shifts.  

The study of the polarized differential cross-section difference
revealed that the reduction by inclusion of $NN$-rescattering appears
predominantly at pion forward angles by about 15 percent. For pions
emitted in the backward direction the $NN$-rescattering effect is 
completely negligible. In comparison with experiment, a quite satisfactory
agreement is obtained.  The polarized total 
cross sections for  circularly polarized photons on a target with spin
parallel $\sigma^P$ and antiparallel $\sigma^A$ to the photon
spin are also investigated. The contribution of $\vec\gamma\vec
d\to\pi^-pp$ to the spin 
response of the deuteron has been explicitly evaluated over the range
of the $\Delta$(1232)-resonance with inclusion of $NN$-rescattering.
In the case of $\sigma^P$, we obtained a significant difference
between the IA and the elementary one, whereas for $\sigma^A$ the
difference is smaller. We found that $NN$-FSI effect appears mainly in $\sigma^P$.  It leads to a strong reduction
of the spin asymmetry in the energy region of the
$\Delta$(1232)-resonance. This reduction becomes about 35 $\mu$b in
the maximum. For the unpolarized total cross section, we found that
$NN$-rescattering  reduces the total cross section in the
$\Delta$(1232)-resonance region by about 5 percent. In comparison with
experiment, the inclusion of such effect leads to an improved
agreement with experimental data.  

It remains as a task for further theoretical research to investigate
the reaction $\gamma d\to\pi NN$ including a three-body treatment in 
the final $\pi NN$ system. This extension is desirable for the
calculation of such rescattering for further developments. Instead of
a separable potential, a more realistic potential for the
$NN$-scattering should be considered. A further interesting topic
concerns the study of polarization observables with the inclusion of
rescattering effects. These studies give more detailed 
information on the $\pi NN$ dynamics and thus providing more stringent
tests for theoretical models. As future refinements we consider also
the use of a more sophisticated elementary production operator, which
will allow one to extend the present results to higher 
energies. A measurement of the spin asymmetry for the deuteron is
needed.
\begin{ack}
I am gratefully acknowledge very useful discussions with Prof.\ H.\ 
Arenh\"ovel as well as the members of his work group. I would like to 
thank the members of the GDH collaboration, specially Dr.\ P.\ Pedroni, 
for providing us with their experimental data. 
\end{ack}

\end{document}